\documentstyle[12pt,catmac]{article}

\def\boxit#1#2{\setbox1=\hbox{\kern#1{#2}\kern#1}%
\dimen1=\ht1 \advance\dimen1 by #1 \dimen2=\dp1 \advance\dimen2 by #1
\setbox1=\hbox{\vrule height\dimen1 depth\dimen2\box1\vrule}%
\setbox1=\vbox{\hrule\box1\hrule}%
\advance\dimen1 by .4pt \ht1=\dimen1
\advance\dimen2 by .4pt \dp1=\dimen2 \box1\relax}

\def\bo#1{\boxit{1pt}{$#1$}}

\font\petit=cmti7 scaled 1200

\def\fsf{{\bf Q}{\not<}\bb{A}{\not>}}

\def\vinv{\vrule height 4mm depth 0mm width 0mm}
\def\vvinv#1{\vrule height 0mm depth #1mm width 0mm}

\font\goth=eufb10 scaled 1200

\newfont{\bb}{msbm10}

\newtheorem{example}{Example}[section]

\newtheorem{theorem}[example]{Theorem}

\newtheorem{definition}[example]{Definition}
\newtheorem{proposition}[example]{Proposition}

\newtheorem{lemma}[example]{Lemma}

\def\boxit#1#2{\setbox1=\hbox{\kern#1{#2}\kern#1}%
\dimen1=\ht1 \advance\dimen1 by #1 \dimen2=\dp1 \advance\dimen2 by #1
\setbox1=\hbox{\vrule height\dimen1 depth\dimen2\box1\vrule}%
\setbox1=\vbox{\hrule\box1\hrule}%
\advance\dimen1 by .4pt \ht1=\dimen1
\advance\dimen2 by .4pt \dp1=\dimen2 \box1\relax}
\def\bo#1{\boxit{1pt}{$#1$}}

\def\Proof{\medskip\noindent {\it Proof --- \ }}
\def\cqfd{\hfill $\Box$ \bigskip}
\def\adots{\mathinner{\mkern2mu\raise1pt\hbox{.}
\mkern3mu\raise4pt\hbox{.}\mkern1mu\raise7pt\hbox{.}}}
\def\<{\langle}
\def\>{\rangle}

\def\det{{\rm det}}

\def\Q{\hbox{\bf Q}}

\def\QA{\Q{<} \bb{A}{>}}
\def\fsfab{\Q{\not<}  a,b{\not>}}

\title{MINOR IDENTITIES FOR \\ QUASI-DETERMINANTS  \\
AND QUANTUM DETERMINANTS}

\author{Daniel KROB
\thanks{\ Institut Blaise Pascal (LITP) -- Universit\'e Paris 7 --
2, place Jussieu -- 75251 Paris Cedex 05 -- France.
E-mail: dk@litp.ibp.fr}
\ and \
Bernard LECLERC
\thanks{\ Institut Gaspard Monge -- Universit\'e de Marne-la-Vall\'ee -
93160 Noisy-le-Grand -- France. E-mail: bl@litp.ibp.fr}
}
\date{\small July 1994}

\begin{document}

\newfont{\rml}{cmu10}
\newfont{\rmL}{cmr12}
\maketitle

\bigskip\bigskip
\bigskip\bigskip
\bigskip

\begin{abstract}
{\rm  We present several identities involving quasi-minors of
noncommutative generic matrices. These iden\-tities are specialized
to quantum matrices, yielding {$q$}-analogues of various classical
determinantal for\-mu\-las.}
\end{abstract}

\vskip 17mm


\section{Introduction}

A common feature of the algebraic constructions which originated from the
quantum inverse scattering method is the systematic use of matrices $T$ with
noncommutative entries, obeying a relation of the form
$$ R\,T_1\,T_2 = T_2\,T_1\,R $$
where the $R$-matrix is a solution of the Yang-Baxter equation
\cite{Dr1,Ji1,RTF}.
The entries of the monodromy matrix $T$ may be regarded as the generators
of an associative algebra subject to the above relation.
Many interesting examples of algebras arise in this way. Among them are
$A_q(GL_n)$, the quantized algebra of functions
on $GL_n$ \cite{RTF}, the quantized universal
enveloping algebra $U_q(gl_n)$ \cite{Dr1,Ji1,RTF}, the Yangian $Y(gl_n)$
\cite{Dr1,O,MNO} and the quantized Yangian $Y_q(gl_n)$ \cite{Ch}. In each of
these cases,
an appropriate concept of {\it quantum determinant}
can be defined \cite{KS,Ji2,RTF} which is
of fundamental importance in the description of the center of these algebras
and
their representation theory. For example the Drinfeld generators \cite{Dr2} of
the Yangian
$Y(gl_n)$ are given by some {\it quantum minors} of the $T$-matrix.
These generators can be used to construct the Gelfand-Zetlin bases for certain
irreducible representations of $Y(gl_n)$ \cite{NT,Mo}.
Moreover, it is shown in \cite{NT} that the Gelfand-Zetlin formulas
for $U_q(gl_n)$ follow from certain algebraic identities satisfied by quantum
minors
of the $T$-matrix corresponding to the quantized Yangian $Y_q(gl_n)$.
Another application of quantum determinants is the construction of a
$q$-deformation
of the coordinate ring of the Grassmanian and the flag manifold, whose basis
consists in products of quantum minors of the $T$-matrix associated with the
algebra
$A_q[GL_n]$ \cite{LR,TT}. In this case, the quadratic relations satisfied by
quantum minors
can be used to establish an analogue of the classical straightening formula
\cite{CLL}.
These examples suggest that it is an important
task to explore these various quantum determinants and to investigate the
algebraic
relations between their minors.

In fact, the problem of defining the determinant of a matrix with
noncommutative
entries is an old one and can be traced back to Cayley \cite{Ca}. An example of
great significance in the classical representation theory is Capelli's
determinant
\cite{Cap,We,Ho,N,NUW}. In the forties, Dieudonn\'e
proposed his famous definition of the determinant of a matrix over a
noncommutative skew-field
\cite{Di} which was subsequently used and extended by Sato and Kashiwara in the
context of the theory of pseudo-differential operators
\cite{SaKa}. Another
interesting construction is that of Berezin who defined an analogue of the
determinant for supermatrices \cite{Be}. However, it is only recently that
Gelfand
and Retakh initiated a completely different approach, introducing the {\it
quasi-determinants}
of a matrix with noncommutative entries \cite{GR1,GR2}.

The most striking facts about quasi-determinants are the following: (1) a $n
{\times} n$ matrix
$A=(a_{ij})$ admits not only one but (in general) $n^2$ quasi-determinants
related by the so-called
homological relations; (2) the quasi-determinants of $A$ are not polynomials
but noncommutative rational
functions of the $a_{ij}$;
(3) in contrast to the Capelli determinant or
to the various quantum determinants which only make sense for very particular
matrices
with entries
obeying some specific commutation rules, quasi-determinants are defined for
matrices
with formal noncommutative entries, and can therefore be specialized to any
matrix;
(4) the Capelli determinant, the Dieudonn\'e determinant, the Berezin
determinant and the quantum determinants of $A_q[GL_n]$
and $Y(gl_n)$ can be expressed in a uniform
way as products of commuting quasi-determinants.

The aim of this article is to demonstrate that the quasi-determinants of
Gelfand and
Retakh can be applied successfully to the important problem of describing the
algebraic relations satisfied by the quantum minors of a monodromy matrix. To
this end,
we first investigate identities satisfied by
{\it quasi-minors} of the generic noncommutative matrix, and then derive
quantum determinantal identities by specializing them to a $T$-matrix.
For simplicity, we only consider in this paper the $T$-matrix of the generators
of the quantum group $A_q(GL_n)$, but the same technique applies also to
the case of $Y(gl_n)$.

We emphasize that from our point of view, the generic quasi-minors identities
are perhaps more important than their specializations to a given monodromy
matrix.
Indeed, they lend themselves to other applications, as illustrated
by \cite{GKLLTR} where the same identities are used for studying noncommutative
symmetric functions, Pad\'e approximants and orthogonal polynomials.
Noncommutative Pad\'e approximants and orthogonal polynomials
appear for instance in Quantum Field theory where they are used for computing
rational
approximations of pertubation series \cite{Bes,GG}.

The paper is organized as follows.
Section 2.1 introduces the {\it free field}, which is the natural algebraic
setting for dealing with quasi-determinants. Section 2.2  provides
a self-contained introduction to quasi-determinants and their basic properties.
It happens that
quasi-determinants are closely related to the representation aspect
of automata theory initiated by Sch\"utzenberger (see \cite{BR}, \cite{Sc}).
The presentation we give here is influenced by this point of view.
We describe then in Section 2.3
noncommutative analogues of several classical theorems, including
Jacobi's theorem, Cayley's
law of complementaries, Muir's law of extensible minors, Sylvester's
theorem, Bazin's theorem and Schweins' series. Finally, these results are
specialized in Section 3 to quantum minors of $A_q(GL_n)$, yielding quantum
analogues of
the same theorems.


\section{Quasi-determinants}

\subsection{The free field}

Let $A$ be a set of noncommutative indeterminates. We denote by
$\QA$ the free associative algebra generated by $A$ over $\Q$. We wish
to imbed $\QA$ in a field, called its universal field of fractions, or
{\it free field}. In other words, the problem is to extend to noncommutative
polynomials the classical construction of the field of fractions of
a ring of commutative polynomials. There are different equivalent
definitions of the free field due to Amitsur, Bergman and Cohn.
Cohn's approach, which relies on the resolution of linear systems
with coefficients in $\QA$, is the most closely related to
the definition of quasi-determinants \cite{Co1,Co2}.
We shall recall his construction in the case where $\Q$ is the ground field,
the general case
(where $\Q$ is replaced by an arbitrary field) being essentially the same.

{}From a categorical viewpoint, the problem may be formulated as follows.
A $\QA$-{\it field} is a (skew) field $K$ equiped with some ring morphism
$\varphi_K$ from $\QA$ into $K$ such that $K$ is the least field containing
the image of $\varphi_K$. A {\it specialization} between two $\QA$-fields
$K,\,L$ is
a ring homomorphism $\varphi$ from a subring $K_0$ of $K$ to $L$ such that
any element of $K_0$ which is not in the kernel of $\varphi$ has an inverse
in $K_0$.
The class of fields does not form a category, for it is not
possible to define a notion of morphism of fields due to the zero inverting
problem.
However,
one can show that $\QA$-fields equiped with specializations form
a category. Moreover, there is an initial object in this
category which is exactly the so-called free field $\fsf$.

\begin{center}
\resetparms
\qtriangle[\QA`\fsf`K;i`\varphi_K`\overline{\varphi_K}]
\end{center}

\noindent
In other words, for every $\QA$-field $K$, there is a unique specialization
$\overline{\varphi_K}$ from $\fsf$ to $K$ that extends $\varphi_K$.

More concretely, here is how Cohn constructs $\fsf$. A $n\times n$ matrix $M$
with entries in $\QA$ is called a {\it full} matrix if it cannot be
written as a product of an $n\times r$ by an $r\times n$ matrix where $r<n$.
$M$ is called {\it linear} if its entries have degree $\le 1$. Let $\Sigma$
be the set of full linear matrices, and for each $n\times n$ matrix
$M=(m_{ij})$
in $\Sigma$, take a set of $n^2$ symbols, arranged as an $n\times n$ matrix
$M'=(m'_{ij})$. Define a ring by the presentation consisting of all the
elements of $\QA$, as well as all the $m'_{ij}$ as generators, and as
defining relations take all the relations
$$
M\,M'=M'\,M=I_n
$$
for each $M$ in $\Sigma$. This ring is none other than the free field $\fsf$.
Using this construction one can show that
any element $x$ of $\fsf$ can be represented as
\begin{equation} \label{MINREP}
x = I \, M^{-1} \, T \ ,
\end{equation}
where $I = ( 1,\,  0, \, \dots , \, 0)$ considered as an $n$-dimensional
row vector, $T$ is some column vector of $\Q^n$ and $M$ is some
element of $\Sigma$. This means that every element $x$ of $\fsf$ is the first
component
of the solution $X$ of some linear system of the form $M\, X = T$.
Cohn and Reutenauer have recently shown the unicity (up to linear isomorphisms)
of the representation of an element $x \in \fsf$ under the form (\ref{MINREP})
when the dimension $n$ is minimal \cite{CR}.

\medskip
There is another interesting construction of the free field based on
Malcev-Neumann series. This method provides a series expansion for every
element of $\fsf$. We first recall a general construction related
to ordered groups introduced independently by Malcev and Neumann
\cite{Mal,Ne}.

Let $\le$ be some total order on a group $G$ compatible with
the group structure, which means that
$$
g_1 \le h_1 \ , \ g_2 \le h_2 \ \ \Longrightarrow \ \ g_1\, g_2 \le h_1 \, h_2
$$
for any $g_1,g_2,h_1,h_2 \in G$. A {\it Malcev-Neumann series} is
a formal series over $G$ whose support is well-ordered with respect to $\le$.
Malcev-Neumann series can therefore be multiplied, and one can show
that they form a field denoted by $\Q_M[[G]]$.

Consider now the free group $F(A)$ constructed over $A$. There are several
classical ways of totally ordering $F(A)$, based on the fact that the
successive quotients of the lower central series of $F(A)$ are free
abelian groups \cite{Pa}. Hence one can
consider the Malcev-Neumann series field $\Q_M[[F(A)]]$.
One can show that the subfield of $\Q_M[[F(A)]]$ generated by the
group algebra $\Q[F(A)]$ is always (independently of the order $\le$ chosen
on $F(A)$) isomorphic to $\fsf$.

Consider for instance the element $(ab-ba)^{-1}$ of $\fsfab$.
Choose an order on $F(a,b)$ such that $ba \le ab$. Then
$$
1\ge bab^{-1}a^{-1} \ge (bab^{-1}a^{-1})^2 \ge \cdots \,,
$$
and the ex\-pan\-sion of $(ab-ba)^{-1}$ as a Malcev-Neumann series is
$$
(ab-ba)^{-1}
=
(ab)^{-1}\, (1-b\, a b^{-1} a^{-1})^{-1}
=
(ab)^{-1} \ \sum_{n=0}^{+\infty} \ (b\, a b^{-1} a^{-1})^n\,.
$$


\subsection{Definition of quasi-determinants} \label{sec:2.1}

We let now
$\bb{A} = \{$ $a_{ij},\, 1 \leq i,j \leq n\, \}$ be an alphabet of
$n^2$ letters. The matrix
$( a_{ij} )_{1\leq i,j\leq n}$, also denoted by $A$, is called the {\it generic
matrix}
of order $n$. It is a full linear matrix, as well as all its submatrices.
Therefore, all square submatrices of $A$ are invertible in $\fsf$.
Throughout the paper we shall use the following notation for submatrices.
For $P$, $Q$ subsets of $\{1,\,\ldots ,\,n\}$, we let $A_{PQ}$ denote
the submatrix whose row indices belong to $P$ and column indices to $Q$,
and $A^{PQ} = A_{\overline{P}\,\overline{Q}}$, where
$\overline{P}=\{1,\,\ldots,\,n\}\, \backslash \, P$ and
$\overline{Q}=\{1,\,\ldots,\,n\}\, \backslash \, Q$ are the set complements of
$P$ and $Q$.
Consider a block decomposition of $A$
$$
A= \matrix{\hbox{\petit P} \cr \hbox{\petit Q}}
\vbox{
\hbox{\hskip5.4mm \hbox{\petit R} \hskip 11mm \hbox{\petit S} \vvinv{1}}
\hbox{$
\left(\
\matrix{
A_{PR}& A_{PS} \cr
A_{QR}& A_{QS} \cr
}
\ \right)
$}
}
\,,
$$
and the corresponding decomposition of $A^{-1}= B = (b_{ij})$
$$
A^{-1}= B = \matrix{\hbox{\petit R} \cr \hbox{\petit S}}
\vbox{
\hbox{\hskip5.4mm \hbox{\petit P} \hskip 11mm \hbox{\petit Q} \vvinv{1}}
\hbox{$
\left(\
\matrix{
B_{RP}& B_{RQ} \cr
B_{SP}& B_{SQ} \cr
}
\ \right)
$}
}
\,.
$$
Here we suppose that $|P|=|R|$ and $|Q|=|S|$ so that
$A_{PR},\,A_{QS},\,B_{RP},\,B_{SQ}$ are square matrices.
By block multiplication we get the classical relations
\begin{eqnarray}\label{BLOC}
B_{RP}& = & (A_{PR} - A_{PS} A_{QS}^{-1} A_{QR})^{-1} \\
B_{RQ}& = & - A_{PR}^{-1} A_{PS} (A_{QS} - A_{QR} A_{PR}^{-1} A_{PS})^{-1}
\label{BLOC2}\\
B_{SP}& = & - A_{QS}^{-1} A_{QR} (A_{PR} - A_{PS} A_{QS}^{-1} A_{QR})^{-1}
\label{BLOC3}\\
B_{SQ}& = & ( A_{QS} - A_{QR} A_{PR}^{-1} A_{PS})^{-1} \label{BLOC4}
\end{eqnarray}
In particular, taking $P=\{p\}$, $R=\{r\}$, one obtains that the entries of the
inverse of $A$
are given by the recursive formula
\begin{equation}\label{INVENTR}
b_{rp}= (a_{pr} - A_{pS} (A_{QS})^{-1} A_{Qr})^{-1} \,,
\end{equation}
where $Q=\{1,\,\ldots ,\,n\} \backslash \{p\}$ and
$S=\{1,\,\ldots,\,n\} \backslash \{r\}$.
This leads to the following definition.

\begin{definition}\label{def:2.3}
{\rm(Gelfand, Retakh; \cite{GR1})$\,$}

Let $A^{pq}$ denote the matrix obtained from $A$ by deleting the
$p$-th row and the $q$-th column. Let also
$\xi_{pq}=(a_{p1},\ldots,\hat a_{pq},\ldots ,a_{pn})$  and
$\eta_{pq}=(a_{1q},\ldots,\hat a_{pq},\ldots ,a_{nq})$.
The quasi-determinant $|A|_{pq}$ of index $pq$
of the generic matrix $A$ is the element of $\fsf$ defined by
\begin{equation}\label{DEF1}
|A|_{pq}= a_{pq} - \xi_{pq}\, (A^{pq})^{-1}\, \eta_{pq}\ =
a_{pq}- \displaystyle\sum_{i\not=p, j\not=q} \
a_{pj}\,((A^{pq})^{-1})_{ji} \, a_{iq} \ ,
\end{equation}
where $\xi_{pq}$ is considered as a row vector and $\eta_{pq}$ as a
column vector.
\end{definition}
It is sometimes convenient to use the following more explicit notation
$$
|A|_{pq}= \left|\matrix{
a_{11}& \ldots & a_{1q} & \ldots & a_{1n} \cr
\vdots&        & \vdots &        & \vdots \cr
a_{p1}& \ldots & \bo{a_{pq}} & \ldots & a_{pn} \cr
\vdots&        & \vdots &        & \vdots \cr
a_{n1}& \ldots & a_{nq} & \ldots & a_{nn} \cr
}\right|\ .
$$
For instance, for $n = 2$ there are four quasi-determinants
$$
\left|
\matrix{
\bo{a_{11}}& a_{12} \cr
a_{21}& a_{22} \cr
}
\right|
=a_{11} - a_{12}\, a_{22}^{-1}\, a_{21}\ , \ \
\left|
\matrix{
a_{11}& \bo{a_{12}} \cr
a_{21}& a_{22} \cr
}
\right|
=a_{12} - a_{11}\, a_{21}^{-1}\, a_{22}\ ,
$$
$$
\left|
\matrix{
a_{11}& a_{12} \cr
\bo{a_{21}}& a_{22} \cr
}
\right|
=a_{21} - a_{22}\, a_{12}^{-1}\, a_{11}\ , \ \
\left|
\matrix{
a_{11}& a_{12} \cr
a_{21}& \bo{a_{22}} \cr
}
\right|
=a_{22} - a_{21}\, a_{11}^{-1}\, a_{12}\ .
$$

\smallskip
The quasi-determinants $|M|_{pq}$ of a matrix $M=(m_{ij})$ with entries in
a given field $K$ are obtained by applying the specialization
$a_{ij}\longrightarrow m_{ij}$ to the rational expressions $|A|_{pq}$.
Some of them may fail to be defined. A sufficient condition
for $|M|_{pq}$ to be well-defined is that $M^{pq}$ is invertible in $K$.
It follows from (\ref{INVENTR}) and (\ref{DEF1}) that  when $K$ is a
commutative field,
$|M|_{pq}= (-1)^{p+q}\, {\rm det}M/{\rm det}M^{pq}$.
Thus quasi-determinants may be regarded as noncommutative analogues
of the ratio of a determinant to one of its principal minors.

By construction the quasi-determinants of the generic matrix $A$ are the
inverses
of the entries of $B = A^{-1}$:
\begin{equation}\label{INV}
b_{ij}^{-1}= |A|_{ji},\ \ i,\,j = 1,\,\ldots ,\,n\,.
\end{equation}
Thus we can rewrite (\ref{DEF1}) as
\begin{equation}\label{eq:2.3}
|A|_{pq}=
a_{pq}- \displaystyle\sum_{i\not=p, j\not=q} \
a_{pj}\,(|A^{pq}|_{ij})^{-1} \, a_{iq} \, ,
\end{equation}
which can be regarded as a recursive definition of $|A|_{pq}$.

We now recall from \cite{GR1,GR2} how quasi-determinants behave under
elementary operations on rows and columns.
\begin{proposition}\label{prop:2.9}
A permutation of the rows or columns of a quasi-determinant does not
change its value.
\end{proposition}

\Proof Let $\sigma \in \goth{S}_n$ and let $P_\sigma$ be the associated
permutation matrix. Then we have
$$
|P_\sigma\,A\, P_\sigma^{-1}|^{-1}_{pq} =
((P_\sigma\,A\, P_\sigma^{-1})^{-1})_{qp} =
((P_\sigma\,A^{-1}\, P_\sigma^{-1})_{qp} = (A^{-1})_{\sigma(q)\sigma(p)} =
|A|^{-1}_{\sigma(p)\sigma(q)}
$$
\hbox{} \cqfd

For example,
$$
\left|\matrix{
a_{11}& a_{12} & a_{13} \cr
a_{21}& a_{22} & a_{23} \cr
\bo{a_{31}}& a_{32} & a_{33} \cr
}\right|
=
\left|\matrix{
a_{21}& a_{22} & a_{23} \cr
a_{11}& a_{12} & a_{13} \cr
\bo{a_{31}}& a_{32} & a_{33} \cr
}\right|
=
\left|\matrix{
a_{22}&a_{21} & a_{23} \cr
a_{12}&a_{11} & a_{13} \cr
a_{32}&\bo{a_{31}} & a_{33} \cr
}\right|\,.
$$
\smallskip
\begin{proposition}\label{prop:2.11}
If the matrix $C$ is obtained from the matrix $A$ by multiplying the $p$-th
row {\rm on the left} by $\lambda$, then
$$
|C|_{kq}=
\left\{\matrix{
\lambda\,|A|_{pq} & {\rm for} \ \ k=p \ , \cr
|A|_{kq} & {\rm for} \ \ k\not =p \ .\cr
}\right.
$$
Similarly, if the matrix $C$ is obtained from the matrix $A$ by multiplying
the $q$-th column {\rm on the right} by $\mu$, then
$$
|C|_{pl}=
\left\{\matrix{
|A|_{pq}\,\mu & {\rm for} \ \ l=q\ , \cr
|A|_{pl} & {\rm for} \ \ l\not =q\ .\cr
}\right.
$$
Finally, if the matrix $D$ is obtained from $A$ by adding to some row
(resp.column) of $A$ its $k$-th row (resp. column), then
$|D|_{pq}= |A|_{pq}$ for every $p \not=k$ (resp. $q \not= k$).
\end{proposition}

\Proof The two first properties follow from (\ref{eq:2.3}) by induction on $n$.
Let $D$ be obtained from $A$ by adding its $k$-th row to its
$l$-th row, and set $M = I_n + E_{lk}$, where $E_{lk}$ denotes the matrix
whose unique non-zero entry is the $lk$-th entry equal to $1$.
Then $D = M A$, and
$$
|D|^{-1}_{pq}= (D^{-1})_{qp} = (A^{-1} M^{-1})_{qp} = (A^{-1})_{qp}
=|A|^{-1}_{pq}
$$
for every $p \not= k$, since multiplying a matrix by $M$ on the right
modifies only its $k$-th column. \cqfd

A major difference between quasi-determinants and determinants is that
quasi-deter\-mi\-nants are not polynomials but rational functions of
the entries of the matrix.
However, formal power series expansions of quasi-determinants can be
obtained, which are conveniently described in terms of graphs.
To this end, we introduce the field
automorphism $\omega$ defined by setting
\medskip
\vskip 0.5mm
\centerline{$
\omega(a_{ij}) = \left\{ \
\matrix{
1 - a_{ii} & \hbox{if} \ \ i = j \cr
- a_{ij} & \hbox{if} \ \ i \not= j \cr
}
\right.
$}

\medskip
\vskip 0.5mm
\noindent
for $1\leq i,j\leq n$ \cite{Co2}. This involution maps the generic matrix $A$
on $I - A$,
and its inverse on the {\it star} of the matrix $A$
$$
A^* = (I-A)^{-1} = \sum_{i=0}^{+\infty} \ A^i \ .
$$
The star operation is a basic tool of automata theory \cite{BR}, and $\omega$
establishes
a correspondence between quasi-deter\-mi\-nants and formal power series
associated
with automatas. In our case, it is useful to associate with $A$ the automaton
${\cal A}$
whose transition matrix is $A$. In other words, ${\cal A}$
is the complete oriented graph constructed over $\{ 1,\dots,n \}$,
the edge from $i$ to $j$ being labelled by $a_{ij}$.
Thus, for $n = 2$, the automaton ${\cal A}$ is

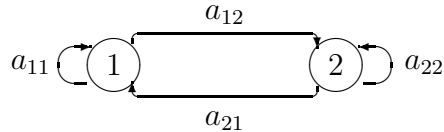
\setlength{\unitlength}{0.7pt}

\centerline{
\begin{picture}(170,92)(-25,-40)
  \put(0,0){\circle{29}}
  \put(0,0){\makebox(0,0){{\rm 1}}}
  \put(120,0){\circle{29}}
  \put(120,0){\makebox(0,0){{\rm 2}}}
  \put(60,12){\oval(100,10)[t]}
  \put(60,27){\makebox(0,0){$a_{12}$}}
  \put(60,-12){\oval(100,10)[b]}
  \put(60,-30){\makebox(0,0){$a_{21}$}}
  \put(-15,0){\oval(30,20)[l]}
  \put(-45,0){\makebox(0,0){$a_{11}$}}
  \put(135,0){\oval(30,20)[r]}
  \put(168,0){\makebox(0,0){$a_{22}$}}
  \put(-13,10){\vector(1,0){1}}
  \put(133,10){\vector(-1,0){1}}
  \put(10,-10){\vector(0,1){1}}
  \put(110,10){\vector(0,-1){1}}
\end{picture}
}

\noindent
Denote by ${\cal P}_{ij}$ the set of words labelling a path in
${\cal A}$ going from $i$ to $j$, {\it i.e.} the set of words of the
form $w = a_{ik_1}\, a_{k_1k_2}\, a_{k_2k_3}\, \ldots\, a_{k_{r-1}j}$.
A {\it simple path} is a path such that $k_s \not = i,\ j$ for every $s$.
We denote by ${\cal SP}_{ij}$ the set of words labelling simple paths from
$i$ to $j$.
It is clear that the entry of index $ij$ of $A^*$ is equal to
$$
(A^{*})_{ij} = \sum_{w\in {\cal P}_{ij}} \, w\,,
$$
or equivalently,
$$
|I-A|_{ij}^{-1} =  \sum_{w \in {\cal P}_{ji}} \ w  \ \ .
$$
Using natural decompositions of these sets of paths, we arrive at the classical
formula
\begin{equation} \label{AUTOMATE}
\left(\,
\matrix{
a_{11} & a_{12} \cr
a_{21} & a_{22}
}
\,\right)^*
=
\left(\,
\matrix{
(a_{11} + a_{12}\, a_{22}^*\, a_{21})^* &
a_{11}^*\, a_{12}\, (a_{22} + a_{21}\, a_{11}^*\, a_{12})^* \cr
a_{22}^*\, a_{21}\, (a_{11} + a_{12}\, a_{22}^*\, a_{21})^* &
(a_{22} + a_{21}\, a_{11}^*\, a_{12})^*
}
\,\right) \,,
\end{equation}
where the star of a series $s$ in the $a_{ij}$ with zero constant
coefficient is defined by
$$s^* = \, \sum_{n\ge 0}\, s^n\,.$$
For instance, the equality of the entries of index $11$ in (\ref{AUTOMATE})
amounts to the decomposition of paths from $1$ to $1$ into sequences of
paths going from $1$ to $1$ without using $1$ as an intermediate state.
We note that (\ref{AUTOMATE}) is to be seen as the image under
$\omega$ of (\ref{BLOC}), (\ref{BLOC2}), (\ref{BLOC3}), (\ref{BLOC4}),
the noncommutative entries $a_{ij}$ of (\ref{AUTOMATE}) being interpreted as
the blocks $A_{PR}$ of some block decomposition of the matrix $A$.
Similarly, one has
\begin{equation}
|I-A|_{ii} =  1-\sum_{{\cal SP}_{ii}} \ w  \ \ .
\end{equation}
For example,
$$
\left|
\matrix{
\bo{1-a_{11}} & -a_{12} \cr
-a_{21} & 1-a_{22} \cr
}
\right|
= 1 - a_{11} -\ \sum_{p\ge 0} \ a_{12}\, a_{22}^p\, a_{21}\ .
$$
The graphical interpretation of formal power series expansions of
quasi-determinants
is an important question. It has been studied at length by Gelfand and Retakh,
and we refer the reader to \cite{GR2} for many other results.


\subsection{Minors identities for quasi-determinants}

In this section, we give noncommutative analogues of several classical
theorems. The reader is referred to \cite{Le} for a review of these
theorems in the commutative case.
We adopt the following convention for indexing {\it quasi-minors}, that is,
quasi-determinants of submatrices.
If $a_{ij}$ is an entry of some submatrix $A^{PQ}$ or $A_{PQ}$, we
denote by $|A^{PQ}|_{ij}$ or $|A_{PQ}|_{ij}$ the
quasi-minor of this submatrix in which $a_{ij}$ is boxed.

\subsubsection{Jacobi's ratio theorem}

In the commutative case, Jacobi's ratio theorem  \cite{Tu,Bo}
states that each minor of the inverse
matrix $A^{-1}$ is equal, up to a sign factor, to the ratio of the
corresponding complementary minor of the transpose of $A$ to ${\rm det}\, A$.
This generalizes the well-known expression of the entries of $A^{-1}$
as ratios of principal minors of $A$ to ${\rm det}\,A$. The corresponding
statement in the noncommutative case is even more natural.

\begin{theorem}\label{th:2.16}
{\rm(Gelfand, Retakh; \cite{GR1}) $\,$}
Let $A$ be the generic matrix of order $n$, let $B$ be its inverse and let
$(\{i\},L,P)$ and $(\{j\},M,Q)$ be two partitions of $\{1,2,\ldots, n\}$
such that $|L| = |M|$ and $|P| = |Q|$. Then there holds :
$$
|B_{M\cup\{j\},L\cup\{i\}}|_{ji}
=
|A_{P\cup\{i\},Q\cup\{j\}}|_{ij}^{-1}
\, .
$$
\end{theorem}

\Proof Using appropriate permutation matrices allows to reduce the proof
to the case $i = j$, $L = M$ and $P = Q$. Set $R = P \cup \{i\}$.
Formula (\ref{BLOC}) yields
$$
(A_{R\,R})^{-1}= B_{R\,R} - B_{R\,L}\,(B_{L\,L})^{-1}\,B_{L\,R} \,.
$$
Now, considering the entry of index $ii$ of the matrices on both sides, we find
$$
|A_{RR}|^{-1}_{ii} = b_{ii} - \sum_{k,l \in L}
b_{ik}\,|B_{LL}|^{-1}_{lk}\,b_{li}
                   = |B_{L\cup\{i\},L\cup\{i\}}|_{ii} \,.
$$ \cqfd

For example,
take $n=5$, $i=3$, $j=4$, $L=\{1,2\}$,
$M=\{1,3\}$, $P=\{4,5\}$ and $Q=\{2,5\}$. Theorem \ref{th:2.16} shows that
$$
\left|\matrix{
a_{32} & \bo{a_{34}} & a_{35} \cr
a_{42} & a_{44} & a_{45} \cr
a_{52} & a_{54} & a_{55} \cr
}\right|
=
\left|\matrix{
b_{11} & b_{12} & b_{13} \cr
b_{31} & b_{32} & b_{33} \cr
b_{41} & b_{42} & \bo{b_{43}} \cr
}\right|^{-1}\ .
$$
%


\subsubsection{Cayley's law of complementaries} \label{sec:2.3.2}

In the commutative case, Cayley's law of complementaries assumes the following
form. Let $I$ be an identity between minors of the generic matrix $A$. If
every minor is replaced by its complement in $A$ (multiplied by a suitable
power of ${\rm det}\, A$), a new identity $I^C$ is obtained, which is said to
result
from $I$ by application of the law of complementaries \cite{Mu,Bo}.
In the noncommutative case, we have the following analogue of this law.

\begin{theorem} \label{th:2.18}
Let $I$ be an identity between quasi-minors of the
generic matrix $A$ of order $n$. If every quasi-minor $|A_{L,M}|_{ij}$
involved in $I$ is replaced by
$|A_{\overline{M}\cup\{j\},\overline{L}\cup\{i\}}|_{ji}^{-1}$, where
$\overline{L}=\{1,\ldots ,n\}\, \backslash \, L$ and
$\overline{M}=\{1,\ldots ,n\}\, \backslash \, M$, there results a new identity
$I^C$.
\end{theorem}

\Proof Applying identity $I$ to $A^{-1}$ gives identity $I^{C}$ by means
of Theorem \ref{th:2.16}. \cqfd

For example, let $n=3$ and let $I$ be the identity:
$$
\hfill
a_{13}^{-1}
\left|\matrix{
\bo{a_{12}} & a_{13} \cr
a_{32} & a_{33} \cr
}\right|
= a_{13}^{-1}a_{12}-a_{33}^{-1}a_{32} \,.
$$
By means of the law of complementaries, one can deduce from $I$
the new identity $I^{C}$ :
$$
\left|\matrix{
a_{11} & a_{12} & a_{13} \cr
a_{21} & a_{22} & a_{23} \cr
\bo{a_{31}} & a_{32} & a_{33} \cr
}\right|
\left|\matrix{
a_{11} & a_{12} \cr
\bo{a_{21}} & a_{22} \cr
}\right|^{-1}
=
\left|\matrix{
a_{11} & a_{12} & a_{13} \cr
a_{21} & a_{22} & a_{23} \cr
\bo{a_{31}} & a_{32} & a_{33} \cr
}\right|
\left|\matrix{
a_{11} & a_{12} & a_{13} \cr
\bo{a_{21}} & a_{22} & a_{23} \cr
a_{31} & a_{32} & a_{33} \cr
}\right|^{-1}
$$
$$
\hfill
-
\left|\matrix{
a_{11} & a_{12} & a_{13} \cr
a_{21} & a_{22} & a_{23} \cr
a_{31} & a_{32} & \bo{a_{33}} \cr
}\right|
\left|\matrix{
a_{11} & a_{12} & a_{13} \cr
a_{21} & a_{22} & \bo{a_{23}} \cr
a_{31} & a_{32} & a_{33} \cr
}\right|^{-1} \,.
$$
%


\subsubsection{Muir's law of extensible minors}

Let us first recall Muir's law of extensible minors in the commutative case
\cite{Mu,Bo}. Let $D$ be a square
matrix of order $n+p$, let $A = D_{P,Q},\ C = D^{P,Q}$ where $P,\ Q$ are
two subsets of $\{1,\ldots ,n+p\}$ of cardinality $n$ and let $I$ be an
identity between minors of $A$. When every minor $|A_{L,M}|$ involved in $I$
is replaced by its {\it extension}
$|D_{L\cup\overline{P},M\cup\overline{Q}}|$ (multiplied by a suitable
power of the {\it pivot} $|C|$ if the obtained identity is not homogeneous),
a new identity $I^{E}$ is obtained, which is called an {\it extensional} of
$I$. A similar rule holds in the noncommutative case.

\begin{theorem} \label{th:2.20}
Let $D$ be the generic matrix of order $n+p$, let
$A = D_{P,Q}$ where $P,\ Q$ are subsets of $\{1,\ldots ,n+p\}$ of cardinality
$n$ and let $I$ be an identity between quasi-minors of $A$. If every
quasi-minor $|A_{L,M}|_{ij}$ involved in $I$ is replaced by its extension
$|D_{L\cup\overline{P},M\cup\overline{Q}}|_{ij}$, a new identity $I^{E}$ is
obtained which is called an extensional of $I$. The submatrix
$D_{\overline{P},\overline{Q}}$ is called the pivot of the extension.
\end{theorem}

\Proof As shown by Muir, Theorem \ref{th:2.20} results from two successive
applications of Theorem \ref{th:2.18}. Indeed, a first application of the
law of complementaries to identity $I$ transforms it into an other identity
$I^{C}$ between quasi-minors of $A$. But quasi-minors of $A$ may be seen as
quasi-minors of $D$ and identity $I^{C}$ may be seen as an identity between
quasi-minors of $D$. A new application to $I^{C}$ of the law of
complementaries, but taking now the complements relatively to $D$,
yields identity $I^{E}$. \cqfd

As an illustration, consider the following identity which results from
(\ref{INV}):
\begin{equation}\label{EXAM}
\hfill
a_{11}
\left|\matrix{
a_{11} & a_{12} & a_{13} \cr
\bo{a_{21}} & a_{22} & a_{23} \cr
a_{31} & a_{32} & a_{33} \cr
}\right|^{-1} \!
+
a_{12}
\left|\matrix{
a_{11} & a_{12} & a_{13} \cr
a_{21} & \bo{a_{22}} & a_{23} \cr
a_{31} & a_{32} & a_{33} \cr
}\right|^{-1} \!
+
a_{13}
\left|\matrix{
a_{11} & a_{12} & a_{13} \cr
a_{21} & a_{22} & \bo{a_{23}} \cr
a_{31} & a_{32} & a_{33} \cr
}\right|^{-1}
= 0 \,.
\end{equation}
An extensional of (\ref{EXAM}) that illustrates Theorem 1.3 of \cite{GR2} is:
\begin{eqnarray*}
\left|\matrix{
\bo{a_{11}}& a_{14} & a_{15} \cr
a_{41}& a_{44} & a_{45} \cr
a_{51}& a_{54} & a_{55} \cr
}\right|
\left|\matrix{
a_{11} & a_{12} & a_{13} & a_{14} & a_{15} \cr
\bo{a_{21}} & a_{22} & a_{23} & a_{24} & a_{25} \cr
a_{31} & a_{32} & a_{33} & a_{34} & a_{35}\cr
a_{41} & a_{42} & a_{43} & a_{44} & a_{45}\cr
a_{51} & a_{52} & a_{53} & a_{54} & a_{55}\cr
}\right|^{-1}
& & \\
+
\left|\matrix{
\bo{a_{12}}& a_{14} & a_{15} \cr
a_{42}& a_{44} & a_{45} \cr
a_{52}& a_{54} & a_{55} \cr
}\right|
\left|\matrix{
a_{11} & a_{12} & a_{13} & a_{14} & a_{15} \cr
a_{21} & \bo{a_{22}} & a_{23} & a_{24} & a_{25} \cr
a_{31} & a_{32} & a_{33} & a_{34} & a_{35}\cr
a_{41} & a_{42} & a_{43} & a_{44} & a_{45}\cr
a_{51} & a_{52} & a_{53} & a_{54} & a_{55}\cr
}\right|^{-1}
& &  \\
+
\left|\matrix{
\bo{a_{13}}& a_{14} & a_{15} \cr
a_{43}& a_{44} & a_{45} \cr
a_{53}& a_{54} & a_{55} \cr
}\right|
\left|\matrix{
a_{11} & a_{12} & a_{13} & a_{14} & a_{15} \cr
a_{21} & a_{22} & \bo{a_{23}} & a_{24} & a_{25} \cr
a_{31} & a_{32} & a_{33} & a_{34} & a_{35}\cr
a_{41} & a_{42} & a_{43} & a_{44} & a_{45}\cr
a_{51} & a_{52} & a_{53} & a_{54} & a_{55}\cr
}\right|^{-1}
& = & 0 \,.
\end{eqnarray*}
Another example is given by the so-called {\it homological relations}
\cite{GR1}.
Start with the trivial identity
$$
a_{il}^{-1}\,
\left|\matrix{
a_{kl} & a_{kj}  \cr
a_{il} & \bo{a_{ij}}\cr
}\right|
=
-a_{kl}^{-1}\,
\left|\matrix{
a_{kl} & \bo{a_{kj}}  \cr
a_{il} & a_{ij}\cr
}\right| \,,
$$
and extend it using the pivot $A^{ki,\,lj}$. The following relation
arises
\begin{equation}
(|A^{kj}|_{il})^{-1} \, |A|_{ij} = - (|A^{ij}|_{kl})^{-1}\, |A|_{kj}\,,
\end{equation}
which relates the two quasi-determinants $ |A|_{ij}$ and $|A|_{kj}$ via
quasi-minors of lower rank. Arguing similarly, one also obtains
\begin{equation}\label{HROW}
|A|_{ij} \, (|A^{il}|_{kj})^{-1} = - |A|_{il}\, (|A^{ij}|_{kl})^{-1}\, .
\end{equation}
The homological relations prove to be a fundamental tool for dealing with
quasi-deter\-mi\-nants. For example, they lead immediately to the following
analogue of the classical expansion of a determinant by a row or column
\begin{equation}\label{ROW}
|A|_{pq} = a_{pq}\, -\
\displaystyle\sum_{j\not = q}\ a_{pj}\,(|A^{pq}|_{kj})^{-1}\, |A^{pj}|_{kq}
\ ,
\end{equation}
\begin{equation}
|A|_{pq} = a_{pq}\, -\
\displaystyle\sum_{i\not = p}\ |A^{iq}|_{pl}\,(|A^{pq}|_{il})^{-1}\, a_{iq}
\ ,
\end{equation}
for any $k \not =p$ and $l \not =q$. Indeed, it follows from (\ref{INV}) that
$$
1 = \ \displaystyle\sum_{j=1}^n \ \, a_{p,j} \, |A|^{-1}_{pj} \ \ .
$$
The row expansion (\ref{ROW}) is obtained by multiplying this last equation
from the right by $|A|_{pq}$, and then using (\ref{HROW}). An explicit
example of (\ref{ROW}), where $n=p=q=4$, is the following
$$
\left|\matrix{
a_{11} & a_{12} & a_{13} & a_{14} \cr
a_{21} & a_{22} & a_{23} & a_{24} \cr
a_{31} & a_{32} & a_{33} & a_{34} \cr
a_{41} & a_{42} & a_{43} & \bo{a_{44}} \cr
}\right|
=
a_{44}-
a_{43}
\left|\matrix{
a_{11} & a_{12} & a_{13} \cr
a_{21} & a_{22} & a_{23} \cr
a_{31} & a_{32} & \bo{a_{33}} \cr
}\right|^{-1}
\left|\matrix{
a_{11} & a_{12} & a_{14} \cr
a_{21} & a_{22} & a_{24} \cr
a_{31} & a_{32} & \bo{a_{34}} \cr
}\right|
$$
$$
-\, a_{42}
\left|\matrix{
a_{11} & a_{12} & a_{13} \cr
a_{21} & a_{22} & a_{23} \cr
a_{31} & \bo{a_{32}} & a_{33} \cr
}\right|^{-1}
\left|\matrix{
a_{11} & a_{13} & a_{14} \cr
a_{21} & a_{23} & a_{24} \cr
a_{31} & a_{33} & \bo{a_{34}} \cr
}\right|
-
a_{41}
\left|\matrix{
a_{11} & a_{12} & a_{13} \cr
a_{21} & a_{22} & a_{23} \cr
\bo{a_{31}} & a_{32} & a_{33} \cr
}\right|^{-1}
\left|\matrix{
a_{12} & a_{13} & a_{14} \cr
a_{22} & a_{23} & a_{24} \cr
a_{32} & a_{33} & \bo{a_{34}} \cr
}\right|
\ .
$$


\subsubsection{Sylvester's theorem}

Another important application of Muir's law of extensible minors is the
noncommuta\-ti\-ve version of Sylvester's theorem. As in the commutative
case, it can be obtained by applying Theorem \ref{th:2.20} to the complete
expansion of a quasi-determinant.

\begin{theorem} \label{th:2.22}
{\rm (Gelfand, Retakh; \cite{GR1}) $\,$} Let
$A$ be the generic matrix of order $n$ and let $P, Q$ be two subsets of
$\{1,\ldots , n\}$ of cardinality $k$. For $i\in\overline{P}$ and
$j\in\overline{Q}$, we set $c_{ij} = |A_{P\cup\{i\},Q\cup\{j\}}|_{ij}$
and form the matrix
$C = (c_{ij})_{i\in\overline{P},j\in\overline{Q}}$ of order $n\!-\! k$.
Then one has
$$
|A|_{lm} = |B|_{lm}
$$
for every $l\in\overline{P}$ and $m\in\overline{Q}$.
\end{theorem}

Let us take $n = 3$, $P = Q =\{3\}$ and
$l = m = 1$. Applying Muir's law to the expansion of
$|A_{\{1,2\},\{1,2\}}|_{11}$ :
$$
\left|\matrix{
\bo{a_{11}} & a_{12} \cr
a_{21} & a_{22} \cr
}\right|
= a_{11} - a_{12}\, a_{22}^{-1}\, a_{21}\ ,
$$
we get the identity
\begin{eqnarray*}
\left|\matrix{
\bo{a_{11}} & a_{12} & a_{13}\cr
a_{21} & a_{22} & a_{23}\cr
a_{31} & a_{32} & a_{33}\cr
}\right|
& = &
\left|\matrix{
\bo{a_{11}} & a_{13} \cr
a_{31} & a_{33} \cr
}\right|
-
\left|\matrix{
\bo{a_{12}} & a_{13} \cr
a_{32} & a_{33} \cr
}\right|
\left|\matrix{
\bo{a_{22}} & a_{23} \cr
a_{32} & a_{33} \cr
}\right|^{-1}
\left|\matrix{
\bo{a_{21}} & a_{23} \cr
a_{31} & a_{33} \cr
}\right|
\\
& = &
\left|\matrix{
{\bo{\left|\matrix{
              \bo{a_{11}} & a_{13} \cr
              a_{31} & a_{33} \cr
              }\right|}}
&
              {\left|\matrix{
              \bo{a_{12}} & a_{13} \cr
              a_{32} & a_{33} \cr
              }\right|}
\cr\noalign{\smallskip}
             {\left|\matrix{
             \bo{a_{22}} & a_{23} \cr
             a_{32} & a_{33} \cr
             }\right|}
&
             {\left|\matrix{
             \bo{a_{21}} & a_{23} \cr
             a_{31} & a_{33} \cr
             }\right|}
\cr
}\right|
\end{eqnarray*}
which is the simplest instance of Sylvester's theorem for
quasi-determinants.

We note that Sylvester's theorem furnishes a recursive method for
evaluating quasi-determinants since it allows to reduce the
computation of a quasi-determinant of order $n$ to the computation of a
quasi-determinant of order $n-1$ whose $(n-1)^2$ entries are quasi-determinants
of order $2$. As one can check, this leads to a cubic
al\-go\-rithm for computing quasi-determinants.

We presented here Sylvester's theorem as a simple consequence of
Theorem \ref{th:2.20}. It can also be directly deduced from relation
(\ref{BLOC}).


\subsubsection{Bazin's theorem} \label{sec:2.3.5}

Sylvester's theorem takes the form of a relation between (quasi-)minors of a
square
matrix. Ba\-zin's theorem deals with maximal (quasi-)minors of a rectangular
matrix. In fact, in both commutative and noncommutative cases, these theorems
are equivalent and each one may be deduced from the other by specialization to
a suitable matrix.

Given a $n\!\times\! 2n$ matrix $A$ and
a subset $P$ of $\{1,\ldots ,2n\}$, we denote for short
by $A_P$ the submatrix $A_{\{1,\ldots n\},P}$.

\begin{theorem} \label{th:2.25}
Let $A$ be the generic matrix of order
$n$ by $2n$. Fix an integer $m$ in $\{1,\ldots,n\}$. For
$1\leq i,j\leq n$, set
$d_{ij} = |A_{\{j,n+1,\ldots ,n+i-1, n+i+1,\ldots ,2n\}}|_{mj}$ and form
the matrix $D = (d_{ij})_{1\leq i,j \leq n}$. Then,
$$
|D|_{kl} = |A_{\{n+1,\ldots ,2n\}}|_{m,n+k} \
|A_{\{1,\ldots,l-1,l+1,\ldots ,n,n+k\}}|_{m,n+k}^{-1} \
|A_{\{1,\ldots ,n\}}|_{ml} \
$$
for any integers $k,l$ in $\{ 1,\ldots, n\}$.
\end{theorem}

\Proof Let us consider the $2n\!\times\! 2n$ matrix $C$ defined by
$$
C = \pmatrix{
A_{\{1,\dots ,n\}} &  A_{\{n+1,\dots, 2n\}} \cr
0_n &  I_n \vinv \cr
}
\ ,
$$
where $I_n$ and $0_n$ denote respectively the unit and zero matrix of
order $n$. Applying Sylvester's theorem to this matrix with
$C_{\{1,\ldots,n\},\{n+1,\ldots,2n\}}$ as pivot, we get
$$
|C|_{n+k,l} = \left|
\matrix{
\left(
\matrix{
\left|
\matrix{
A_j & A_{\{n+1,\ldots,2n\}} \cr
\bo{0} & u_{n+i} \cr
}
\right|
}
\right)_{1\leq i,j \leq n}
}
\right|_{kl}
$$
where $u_i$ denotes for every integer $i$ the row vector whose only non-zero
entry is the $i$-th entry equal to $1$. Expanding by its last row each
quasi-determinant involved in the above identity and using Proposition
\ref{prop:2.11}, we obtain
\begin{equation} \label{eq:2.6}
|C|_{n+k,l} = -\, |A_{\{n+1,\ldots,2n\}}|_{m,n+k}^{-1} \
|D|_{kl} \ .
\end{equation}
On the other hand, it follows from \ref{BLOC2} that
$$
|C|_{n+k,l}^{-1} = - (A_{\{1,\ldots,n\}}^{-1} \,
A_{\{n+1,\ldots,2n\}})_{l,n+k} = -
\displaystyle\sum_{j=1}^n \ |A_{\{1,\ldots,n\}}|_{j,l}^{-1} \, a_{j,n+k} \ .
$$
Using Definition \ref{def:2.3}, this relation can be written in the form
$$
|C|_{n+k,l}^{-1} = \left|
\matrix{
A_{\{1,\ldots,n\}} & A_{n+k} \cr
u_l & \bo{0} \vinv \cr
}
\right| \ .
$$
Expanding now this quasi-determinant by the last row, we get the identity
$$
|C|_{n+k,l}^{-1} = -\, |A_{\{1,\ldots,n\}}|_{ml}^{-1} \
|A_{\{1,\ldots,l-1,l+1,\ldots,n,n+k\}}|_{m,n+k}
$$
and we conclude by comparing to relation (\ref{eq:2.6}).
\cqfd

\begin{example} \label{ex:2.26}
{\rm Let $n = 3$ and $k = l = m =1$. We
adopt more appropriate notations, writing for short $|2\bo{4}5|$ instead
of $|M_{\{2,4,5\}}|_{14}$. Bazin's identity reads}
$$
\left|\matrix{
\bo{|\bo{1}56|} & |\bo{2}56| & |\bo{3}56| \cr
|\bo{1}46| & |\bo{2}46| & |\bo{3}46| \cr
|\bo{1}45| & |\bo{2}45| & |\bo{3}45| \cr
}\right|
=
|\bo{4}56|\ |23\bo{4}|^{-1} \ |\bo{1}23|
\ .
$$
\end{example}


\subsubsection{Schweins' series}

``Schweins found an important series, in 1825, for the quotient of two
$n$-rowed determinants which differ only in one column. This series is of
great use in many branches of algebra and analysis, and many interesting cases
arise by treating one column as a column of the unit matrix" \cite{Tu}.
Here is an example of Schweins' commutative series.
\begin{equation}\label{SCHW}
{(abcd)_{1234}\over (abce)_{1234}}=
{(abc)_{123}(abed)_{1234}\over (abe)_{123}(abce)_{1234}}
+{(ab)_{12}(aed)_{123}\over (ae)_{12}(abe)_{123}}
+{a_1(ed)_{12}\over e_1(ae)_{12}}
+{d_1\over e_1}
\ ,
\end{equation}
where for instance $(aed)_{123}$ denotes the determinant
$
\left|\matrix{
a_1 & e_1 & d_1 \cr
a_2 & e_2 & d_2 \cr
a_3 & e_3 & d_3 \cr
}\right|
$.

\medskip
Schweins' series is still valid in the noncommutative case. Keeping the
notations of \ref{sec:2.3.5}, we first note that according
to the homological
relations one has for a $3\!\times\! 6$ matrix $A$, say,
$$
|A_{123}|_{13}^{-1}\, |A_{124}|_{14} =
|A_{123}|_{23}^{-1}\, |A_{124}|_{24} =
|A_{123}|_{33}^{-1}\, |A_{124}|_{34} \ .
$$
This common value will be denoted for short by $|12\bo{3}|^{-1}|12\bo{4}|$.
We can now state Schweins' series for quasi-determinants. For convenience,
we limit ourselves to the case of quasi-determinants of order 3 and 4, the
general case being easily induced from these.

\begin{theorem} \label{th:2.27}
The maximal quasi-minors of a $3\!\times\! 6$ generic
matrix satisfy the relation
$$
|12\bo{3}|^{-1}|12\bo{4}|=
|12\bo{3}|^{-1}|12\bo{5}|\, |23\bo{5}|^{-1}|23\bo{4}|+
$$
$$
|25\bo{3}|^{-1}|25\bo{6}|\, |35\bo{6}|^{-1}|35\bo{4}|
+|56\bo{3}|^{-1}|56\bo{4}|\ .
$$
The maximal quasi-minors of a $4\!\times\! 8$ generic matrix satisfy the
relation
$$
|123\bo{4}|^{-1}|123\bo{5}|=
|123\bo{4}|^{-1}|123\bo{6}||234\bo{6}|^{-1}|234\bo{5}|
$$
$$
+|236\bo{4}|^{-1}|236\bo{7}||346\bo{7}|^{-1}|346\bo{5}|
+|367\bo{4}|^{-1}|367\bo{8}||467\bo{8}|^{-1}|467\bo{5}|
+|678\bo{4}|^{-1}|678\bo{5}|\ .
$$
\end{theorem}

\Proof Let us take again the notations of Example \ref{ex:2.26}. Applying
Bazin's theorem for $n = 2$ to the matrix $(4513)$, we get :
$$
|\bo{4}1| - |\bo{5}1|\, |\bo{5}3|^{-1} \, |\bo{4}3|
=
\left|
\matrix{
|\bo{4}3| & |\bo{5}3| \cr
\bo{|\bo{4}1|} & |\bo{5}1| \cr
}
\right|
=
|1\bo{3}| \, |5\bo{3}|^{-1} \, |\bo{4}5| \ .
$$
Multiplying from the left by $|1\bo{3}|^{-1}$ and using
Muir's law, one obtains the relation
$$
|12\bo{3}|^{-1}|12\bo{4}|=
|12\bo{3}|^{-1}|12\bo{5}||23\bo{5}|^{-1}|23\bo{4}|
+|25\bo{3}|^{-1}|25\bo{4}|\,.
$$
Schweins' series for order $3$ results
from two applications of this lemma. The general case is similar. \cqfd

As noted by Turnbull, interesting corollaries are obtained by specialization
to a par\-ti\-cu\-lar matrix some columns of which are columns of the unit
matrix. Let us mention the following, which for convenience is stated for
order 3 and 4 only.

\begin{theorem} \label{th:2.28}
The quasi-minors of a $3\!\times\! 4$ generic matrix
satisfy the relation
$$
\left|\matrix{
a_{11} & a_{12} & a_{13}\cr
a_{21} & a_{22} & a_{23}\cr
a_{31} & a_{32} & \bo{a_{33}}\cr
}\right|^{-1}
\left|\matrix{
a_{11} & a_{12} & a_{14}\cr
a_{21} & a_{22} & a_{24}\cr
a_{31} & a_{32} & \bo{a_{34}}\cr
}\right|
=
\left|\matrix{
a_{11} & a_{12} & a_{13}\cr
a_{21} & a_{22} & a_{23}\cr
a_{31} & a_{32} & \bo{a_{33}}\cr
}\right|^{-1}
\left|\matrix{
a_{11} & a_{13} & a_{14}\cr
a_{21} & a_{23} & a_{24}\cr
a_{31} & a_{33} & \bo{a_{34}}\cr
}\right|
$$
$$
+
\left|\matrix{
a_{11} & a_{13}\cr
a_{21} & \bo{a_{23}}\cr
}\right|^{-1}
\left|\matrix{
a_{13} & a_{14}\cr
a_{23} & \bo{a_{24}}\cr
}\right|
+
a_{13}^{-1}a_{14}
\ .
$$
{\it The quasi-minors of a $4\times 5$ generic matrix satisfy the relation}
$$
\left|\matrix{
a_{11} & a_{12} & a_{13} & a_{14}\cr
a_{21} & a_{22} & a_{23} & a_{24}\cr
a_{31} & a_{32} & a_{33} & a_{34}\cr
a_{41} & a_{42} & a_{43} & \bo{a_{44}}\cr
}\right|^{-1}
\left|\matrix{
a_{11} & a_{12} & a_{13} & a_{15}\cr
a_{21} & a_{22} & a_{23} & a_{25}\cr
a_{31} & a_{32} & a_{33} & a_{35}\cr
a_{41} & a_{42} & a_{43} & \bo{a_{45}}\cr
}\right|
$$
$$
=
\left|\matrix{
a_{11} & a_{12} & a_{13} & a_{14}\cr
a_{21} & a_{22} & a_{23} & a_{24}\cr
a_{31} & a_{32} & a_{33} & a_{34}\cr
a_{41} & a_{42} & a_{43} & \bo{a_{44}}\cr
}\right|^{-1}
\left|\matrix{
a_{11} & a_{12} & a_{14} & a_{15}\cr
a_{21} & a_{22} & a_{24} & a_{25}\cr
a_{31} & a_{32} & a_{34} & a_{35}\cr
a_{41} & a_{42} & a_{44} & \bo{a_{45}}\cr
}\right|
+
\left|\matrix{
a_{11} & a_{12} & a_{14}\cr
a_{21} & a_{22} & a_{24}\cr
a_{31} & a_{32} & \bo{a_{34}}\cr
}\right|^{-1}
\left|\matrix{
a_{11} & a_{14} & a_{15}\cr
a_{21} & a_{24} & a_{25}\cr
a_{31} & a_{34} & \bo{a_{35}}\cr
}\right|
$$
$$
+
\left|\matrix{
a_{11} & a_{14}\cr
a_{21} & \bo{a_{24}}\cr
}\right|^{-1}
\left|\matrix{
a_{14} & a_{15}\cr
a_{24} & \bo{a_{25}}\cr
}\right|
+
a_{14}^{-1}a_{15}
\ .
$$
\end{theorem}

\medskip
\Proof Let us prove the first relation, the general case
being similar. We specialize Theorem~\ref{th:2.27} to the $3\!\times\! 6$
matrix
$$
M =
\left(
\matrix{
0 & 0 & a_{13} & a_{14} & a_{11} & a_{12} \cr
1 & 0 & a_{23} & a_{24} & a_{21} & a_{22} \cr
0 & 1 & a_{33} & a_{34} & a_{31} & a_{32} \cr
}
\right) \ .
$$
Using the homological relations (\ref{HROW}), we get
$$
|12\bo{3}|^{-1}\, |12\bo{5}|\, |23\bo{5}|^{-1}\, |23\bo{4}|
=
a_{13}^{-1}\,a_{11}\,
\left|\matrix{
a_{13} & a_{11} \cr
a_{23} & \bo{a_{21}} \cr
}\right|^{-1}
\,
\left|\matrix{
a_{13} & a_{14} \cr
a_{23} & \bo{a_{24}} \cr
}\right|
$$
$$
=
- \left|\matrix{
a_{11} & a_{13} \cr
a_{21} & \bo{a_{23}} \cr
}\right|^{-1}
\,
\left|\matrix{
a_{13} & a_{14} \cr
a_{23} & \bo{a_{24}} \cr
}\right| \ .
$$
Similarly, we also find that
$$
|25\bo{3}|^{-1}|25\bo{6}|\, |35\bo{6}|^{-1}|35\bo{4}|
=
- \left|\matrix{
a_{11} & a_{12} & a_{13}\cr
a_{21} & a_{22} & a_{23}\cr
a_{31} & a_{32} & \bo{a_{33}}\cr
}\right|^{-1}\,
\left|\matrix{
a_{11} & a_{13} & a_{14}\cr
a_{21} & a_{23} & a_{24}\cr
a_{31} & a_{33} & \bo{a_{34}}\cr
}\right| \,,
$$
and the claim follows. \cqfd

Theorem~\ref{th:2.28} is the true analogue for quasi-determinants of
(\ref{SCHW})
and it even looks more natural this way.


\section{Quantum determinants}

In this section we specialize the previous theorems on quasi-determinants
to the matrix of generators of the $q$-deformation $A_q(GL_n)$ of the
algebra of functions on $GL_n$.

\subsection{Definitions and notations}
We recall fundamental facts on the algebras $A_q(Mat_n)$ and $A_q(GL_n)$
\cite{RTF}.
The algebra $A_q(Mat_n)$ is the associative algebra over ${\bf{C}}[q,q^{-1}]$
generated by $n^2$ letters $t_{ij},\, i,\,j = 1,\,\ldots ,\, n$ subject to the
relations
(written in matrix form)
\begin{equation}
R\,T_1\,T_2 = T_2\,T_1\,R \,,
\end{equation}
where
$$
R=q^{-1}\,\sum_{i=1}^n e_{ii}\otimes e_{ii}
+ \sum_{1\le i \not = j \le n} e_{ii}\otimes e_{jj} +
(q^{-1} - q)\, \sum_{1\le i < j \le n} e_{ij}\otimes e_{ji}
\, .
$$
Here, $T=(t_{ij})$, $T_1= T\otimes I$, $T_2= I\otimes T$, and $e_{ij}$'s
are the matrix units. More explicitely, the relations obeyed by the
symbols $t_{ij}$'s can be written
$$
t_{ik}\,t_{il}=q^{-1}\,t_{il}\,t_{ik} \ \ \ {\rm for} \ k<l, \quad
t_{ik}\,t_{jk}=q^{-1}\,t_{jk}\,t_{ik} \ \ \ {\rm for} \ i<j,
$$
$$
t_{il}\,t_{jk} = t_{jk}\,t_{il} \ \ \ {\rm for} \ i<j,\ k<l,
$$
$$
t_{ik}\,t_{jl}-t_{jl}\,t_{ik}= (q^{-1}\!-\! q)\, t_{il}\,t_{jk} \ \ \ {\rm for}
\ i<j,\ k<l\ .
$$
The algebra $A_q(Mat_n)$ has a bialgebra structure whose comultiplication
$\Delta$
and counit $\epsilon$ are given by
$$
\Delta(T) = T\otimes T \ \ \  {\rm i.e.}
\ \ \ \Delta(t_{ij}) = \sum_{k=1}^n t_{ik} \otimes t_{kj}\ \ \  i,\,j =
1,\,\ldots ,\,n\,,
$$
$$
\epsilon(T) = I \ \ \  {\rm i.e.}
\ \ \ \epsilon(t_{ij}) = \delta_{ij}\ \ \  i,\,j = 1,\,\ldots ,\,n\,.
$$
The {\it quantum determinant} of $T$ is the element of $A_q(Mat_n)$ defined by
$$
{\rm det}_q = {\rm det}_q\, T =
\displaystyle\sum_{\sigma\in{\goth{S}}_n} \,
(-q)^{-\ell(\sigma)} \, t_{1{\sigma(1)}}\,\ldots\, t_{n{\sigma(n)}}
\ ,
$$
where ${\goth{S}}_n$ is the symmetric group on $\{1,\ldots ,n\}$ and
$\ell(\sigma)$ denotes the length of the permutation $\sigma$.
The quantum determinant of $T$ belongs to the center of $A_q(Mat_n)$
and is a group-like element, namely $\Delta({\rm det}_q\, T) =
{\rm det}_q\, T \otimes {\rm det}_q\, T$.
More generally, for $P = \{ i_1<\ldots <i_k\}$ and $Q=\{j_1<\ldots <j_k\}$
one defines the {\it quantum minor} of the submatrix $T_{PQ}$ as
$$
{\rm det}_q\, T_{PQ} =
\displaystyle\sum_{\sigma\in{\goth{S}}_k} \,
(-q)^{-\ell(\sigma)} \, t_{i_1j_{\sigma(1)}}\,\ldots\, t_{i_kj_{\sigma(k)}}
\ .
$$
In particular, the {\it quantum comatrix} $C(T) = (c_{ij})$ is defined by
$$
c_{ij} = (-q)^{j-i}\, {\rm det}_q\, (T^{ji})\ \ \  i,\,j = 1,\,\ldots ,\,n\,.
$$
Then, one has
\begin{equation} \label{eq:3.1}
T\,C(T) = C(T)\,T = {\rm det}_q\, T\, .\, I_n \,,
\end{equation}
which amounts to the expansion of $\det_q\, T$ by one of its rows or
columns. This leads to the definition of the algebra $A_q(GL_n)$
as the localization $A_q(Mat_n)[{\rm det}_q^{-1}]$ of $A_q(Mat_n)$.
The algebra $A_q(GL_n)$ is a Hopf algebra whose coproduct and counit
are defined as above, and whose antipode is the anti-automorphism given
by
$$
S(T) = {\rm det}_q^{-1}\,C(T)  \,\ \ \  {\rm i.e.}
\ \ \ S(t_{ij}) = {\rm det}_q^{-1}\,c_{ij}\,,\ \ \ i,\,j = 1,\,\ldots ,\,n\,,
$$
$$
S({\rm det}_q) = {\rm det}_q^{-1} \,.
$$
In other words
\begin{equation}
T\,S(T) = S(T)\,T = I_n \,,
\end{equation}
and $S(T)=(s_{ij})$ is the inverse matrix of $T$.
Finally, we remark that since $S$ is an anti-automorphism, the entries
of $S(T)$ and $C(T)$ obey the same commutation rules as those of $T$
with $q$ replaced by $q^{-1}$.


\subsection{Quantum determinants and quasi-determinants}

We now consider the connection between quantum determinants and
quasi-de\-ter\-mi\-nants. As recalled in Section \ref{sec:2.1},
quasi-determinants are noncommutative analogues of the ratio of a
determinant to one of its principal
minors. Thus if the entries $t_{ij}$ of a matrix $T$ belong to a commutative
field, one has the following expression of ${\rm det}\, T$ in terms of
quasi-determinants
$$
\left|\matrix{
t_{11}& t_{12} &\ldots & t_{1n}\cr
t_{21}& t_{22} &\ldots & t_{2n}\cr
\vdots &\vdots &\ddots & \vdots\cr
t_{n1}& t_{n2} &\ldots & t_{nn}\cr
}\right|
=
\left|\matrix{
\bo{t_{11}}& t_{12} &\ldots & t_{1n}\cr
t_{21}& t_{22} &\ldots & t_{2n}\cr
\vdots &\vdots &\ddots & \vdots\cr
t_{n1}& t_{n2} &\ldots & t_{nn}\cr
}\right|
\left|\matrix{
\bo{t_{22}}& \ldots & t_{2n}\cr
\vdots &\ddots & \vdots\cr
t_{n2}& \ldots & t_{nn}\cr
}\right|
\ \ldots \
\left|\matrix{
\bo{t_{n-1\,n-1}} & t_{n-1\,n}\cr
t_{n\,n-1} & t_{nn}\cr
}\right|
t_{nn}
\ .
$$
The following theorem provides an analogue of this formula for
quantum determinants.

\begin{theorem} \label{th:3.2}
{\rm (Gelfand, Retakh; \cite{GR1}) $\,$} Let
$T = (t_{ij})_{1\leq i,j\leq n}$ be the matrix of generators of $A_q(GL_n)$. In
the field of fractions of $A_q(GL_n)$, one has
$$
{\rm det}_q\, T = |T|_{11} \, |T^{11}|_{22} \,\ldots\, t_{nn}
$$
and the quasi-minors in the right-hand side commute all together. More
generally, let $\sigma=i_1\ldots i_n$ and $\tau=j_1\ldots j_n$ be two
permutations of ${\goth{S}}_n$. There holds
\begin{equation} \label{eq:3.2}
{\rm det}_q\, T=(-q)^{\ell(\sigma)-\ell(\tau)}\,
|T|_{i_1j_1} \, |T^{i_1j_1}|_{i_2j_2}\,\ldots\, t_{i_nj_n}
\end{equation}
and the quasi-minors in the right-hand side commute all together.
\end{theorem}

\Proof We first note that $A_q(GL_n)$ is an Ore ring and therefore has a field
of
fractions.
By (\ref{INV}), the quasi-determinants of
$T$ are the inverses of the entries of
$S(T)=(\det_q\, T)^{-1}\ C(T)$. Hence we have
\begin{equation} \label{eq:3.3}
{\rm det}_q\, T = (-q)^{i-j}\, |T|_{ij}\, \det_q\, T^{ij}
= (-q)^{i-j}\, {\rm det}_q\, T^{ij}\, |T|_{ij} \,,
\end{equation}
and (\ref{eq:3.2}) follows by induction on $n$.
Let us now prove that the quasi-determinants involved in (\ref{eq:3.2})
commute all together. For simplicity, we only argue in the case
$i_1 = j_1 = 1,\, \ldots\, , i_n = j_n = n$. By induction on $n$,
it is enough to prove that $|T|_{11}$ commutes with
$|T^{11}|_{22},\,\ldots\, , t_{nn}$. Using relation (\ref{eq:3.3}), it
suffices to show
that
$\det_q\, T^{11}$ commutes with ${\rm det}_q \,
T^{\{1,\ldots,i\},\{1,\ldots,i\}}$
for $1 \leq i \leq n-1$, which follows from the fact that $\det_qT^{11}$
commutes
with $t_{ij}$, $i,\,j = 2,\,\ldots ,\,n$. \cqfd

Thus, for $n=2$, we have
$$
\det_q\, T = (t_{11}-t_{12}t_{22}^{-1}t_{21}) \, t_{22}
= (-q)^{-1}\, (t_{12}-t_{11}t_{21}^{-1}t_{22})\, t_{21}
$$
$$
= (-q)\, (t_{21}-t_{22}t_{12}^{-1}t_{11})\, t_{12}
= (t_{22}-t_{21}t_{11}^{-1}t_{12})\, t_{11}\ .
$$
Note that the parameter $q$ no longer appears in the first and
fourth expression.


\subsection{Minor identities for quantum determinants}

In this section, we derive quantum analogues of several classical
determinantal formulas. We shall sometimes use the following terminology.
A $k\! \times \!k$ matrix $M=(m_{ij})$ with entries in $A_q(GL_n)$
is called a $k\! \times \!k$ {\it quantum matrix} if its entries $m_{ij}$ obey
the same commutation rules as the generators $t_{ij}$ of $A_q(GL_k)$.
More generally, a rectangular matrix $M$ is said to be a quantum
matrix if all its $2\!\times \!2$ submatrices are quantum matrices.

\subsubsection{Jacobi's ratio theorem}

Recall that Jacobi's theorem states that each minor of the inverse matrix
$A^{-1}$ is equal, up to a sign factor, to the ratio of the corresponding
complementary minor of the transpose of $A$ to $\det\, A$. For quantum
determinants, we have the following analogue.

\begin{theorem} \label{th:3.4}
Let $P = \{i_1<\ldots <i_k\},\ Q=\{j_1<\ldots <j_k\}$ be two subsets of
$\{1,\,\ldots ,\,n\}$, and
$\overline{P}=\{i_{k+1}<\ldots <i_n\}$,
$\overline{Q}=\{j_{k+1}<\ldots <j_n\}$, be their set complements.
Set $\sigma=i_1\ldots i_n$ and
$\tau=j_1\ldots j_n$. Then,
$$
\det_{q^{-1}}\, S(T)_{P,Q}
=
(-q)^{\ell(\tau)-\ell(\sigma)}\, \det_q\, T^{Q,P}\ (\det_q\, T)^{-1} \ .
$$
\end{theorem}

\Proof Express $\det_{q^{-1}} S(T)_{P,Q}$ as a product of
quasi-determinants by means of Theorem \ref{th:3.2} and apply
Jacobi's Theorem \ref{th:2.16} to each of them.
The result then follows from a second application of
Theorem \ref{th:3.2}. \cqfd

For instance, if $n=5$, $P=\{1,3,4\}$ and $Q=\{1,2,3\}$, we have
$$
\left|\matrix{
s_{11} & s_{12} & s_{13} \cr
s_{31} & s_{32} & s_{33} \cr
s_{41} & s_{42} & s_{43} \cr
}\right|_{q^{-1}}
=
(-q)^{-2}
\left|\matrix{
t_{42} & t_{45} \cr
t_{52} & t_{55} \cr
}\right|_q
(\det_q\, T)^{-1} \ .
$$
%

\subsubsection{Cayley's law of complementaries}

For quantum determinants, we obtain the following analogue of Cayley's law
of complementaries (see Section \ref{sec:2.3.2}).

\begin{theorem} \label{th:3.6}
Let $I$ be a polynomial identity with coefficients
in ${\bf{C}}[q,q^{-1}]$ between quantum minors of the
matrix $T$ of generators of $A_q(GL_n)$.
If each minor $\det_q\, T_{P,Q}$ involved in $I$ is replaced by its
complement $\det_q\, T^{P,Q}$ multiplied by $(\det_q\, T)^{-1}$ and if, in
addition, the substitution $q \rightarrow q^{-1}$ is made in the coefficients
of $I$, there results a new identity $I^C$.
\end{theorem}

\Proof The entries of the matrix $S(T)^t$ obey the same commutation rules as
those of $T$ with $q$ replaced by $q^{-1}$. Therefore, replacing in $I$ each
minor $\det_q\, T_{P,Q}$ by $\det_{q^{-1}}\, S(T)^t_{P,Q}$ and subtituting
$q^{-1}$ to $q$ in the coefficients, one obtains a polynomial identity between
quantum minors of $S(T)^t$.
Identity $I^C$ then results from Theorem \ref{th:3.4}. \cqfd

For example, take $n=4$ and consider the following identity (see Proposition
\ref{prop:3.15})
\begin{equation}\label{REL}
\hfill
\left|\matrix{
t_{11} & t_{12}\cr
t_{21} & t_{22}\cr
}\right|_q
\ t_{23}
=
q^{-1} t_{23}
\left|\matrix{
t_{11} & t_{12}\cr
t_{21} & t_{22}\cr
}\right|_q
\,.
\end{equation}
Applying Cayley's law to (\ref{REL}), we get
$$
\hfill
\left|\matrix{
t_{33} &t_{34}\cr
t_{43} &t_{44}\cr
}\right|_q
\left|\matrix{
t_{11} &t_{12} &t_{14}\cr
t_{31} &t_{32} &t_{34}\cr
t_{41} &t_{42} &t_{44}\cr
}\right|_q
=
q\
\left|\matrix{
t_{11} &t_{12} &t_{14}\cr
t_{31} &t_{32} &t_{34}\cr
t_{41} &t_{42} &t_{44}\cr
}\right|_q
\left|\matrix{
t_{33} &t_{34}\cr
t_{43} &t_{44}\cr
}\right|_q
\,.
$$
%


\subsubsection{Muir's law of extensible minors}

{}From Cayley's law \ref{th:3.6} is deduced the following quantum
analogue of Muir's law. The proof is similar to the one of
Theorem~\ref{th:2.20}.

\begin{theorem} \label{th:3.8}
Let $n\le m$ be two integers, and $P,\ Q$ be two
subsets of $\{1,\ldots ,m\}$ of cardinality $n$.
We consider the imbedding of $A_q(GL_n)$ in $A_q(GL_m)$ obtained by
identifying the matrix $T_n$ of generators of $A_q(GL_n)$ to the submatrix
$(T_m)_{PQ}$ of $T_m$. Let  $I$ be a polynomial
identity with coefficients in ${\bf{C}}[q,q^{-1}]$ between quantum minors of
$T_n$.  When every quantum minor $\det_q\,(T_n)_{L,M}$ involved in $I$ is
replaced by
its extension $\det_q\, (T_m)_{L\cup\overline{P},M\cup\overline{Q}}$
(multiplied by
a suitable power of the pivot $\det_q\, (T_m)_{\overline{P},\overline{Q}}$ if
the
identity is not homogeneous), a new identity $I^{E}$ is obtained, which is
called an extensional of $I$.
\end{theorem}

As an illustration, take $n=2,\ m=4,\ P=\{2,4\},\ Q=\{2,3\}$ and
consider the identity
$$
t_{22}\,t_{23}=q^{-1}t_{23}\,t_{22} \,.
$$
Applying Muir's law, we get
$$
\left|\matrix{
t_{11} &t_{12} &t_{14}\cr
t_{21} &t_{22} &t_{24}\cr
t_{31} &t_{32} &t_{34}\cr
}\right|_q
\left|\matrix{
t_{11} &t_{13} &t_{14}\cr
t_{21} &t_{23} &t_{24}\cr
t_{31} &t_{33} &t_{34}\cr
}\right|_q
=
q^{-1}
\left|\matrix{
t_{11} &t_{13} &t_{14}\cr
t_{21} &t_{23} &t_{24}\cr
t_{31} &t_{33} &t_{34}\cr
}\right|_q
\left|\matrix{
t_{11} &t_{12} &t_{14}\cr
t_{21} &t_{22} &t_{24}\cr
t_{31} &t_{32} &t_{34}\cr
}\right|_q
\,.
$$
%


\subsubsection{Sylvester's theorem}

An important consequence of Muir's law is the quantum analogue of Sylvester's
theorem.

\begin{theorem} \label{th:3.10}
Let $\det_q\, T_{P,Q}$ be a quantum $k\!\times\! k$ minor of the matrix
$T$ of generators of $A_q(GL_n)$.
For $i\in\overline{P}$ and $j\in\overline{Q}$, set
$u_{ij}=\det_q\, T_{P\cup\{i\},Q\cup\{j\}}$ and let $U$ denote the
$(n-k)\!\times\! (n-k)$ matrix
$(u_{ij})_{i\in\overline{P},j\in\overline{Q}}$. Then
$U$ is a quantum matrix,
and there holds
$$
\det_q\, U = \det_q\, T\ (\det_q\, T_{P,Q})^{n-k-1}\,.
$$
\end{theorem}

\Proof The commutation rules for the entries of $U$ follow from Muir's law.
Also, applying Muir's law to the complete expansion
of the quantum minor $\det_q\, T^{P,Q}$ yields Sylvester's theorem, the term
$(\det_q\,T_{P,Q})^{n-k-1}$ in the right-hand side being an homogeneity
factor. \hbox{} \cqfd

Thus, let $n=3,\ k=1,\ P=Q=\{3\}$. Sylvester's theorem reads
$$
\left|\matrix{
{\left|\matrix{
t_{11} & t_{13} \cr
                t_{31} & t_{33} \cr
                }\right|_q}
&
                {\left|\matrix{
                t_{12} & t_{13} \cr
                t_{32} & t_{33} \cr
                }\right|_q}
\cr\noalign{\smallskip}
                {\left|\matrix{
                t_{22} & t_{23} \cr
                t_{32} & t_{33} \cr
                }\right|_q}
&
                {\left|\matrix{
                t_{21} & t_{23} \cr
                t_{31} & t_{33} \cr
                }\right|_q}
\cr
}\right|_q
=
\left|\matrix{
t_{11} & t_{12} & t_{13}\cr
t_{21} & t_{22} & t_{23}\cr
t_{31} & t_{32} & t_{33}\cr
}\right|_q
t_{33}\ .
$$

Sylvester's quantum theorem can also be directly
deduced from its noncommutative analogue (Theorem \ref{th:2.22}),
using the same method as in the proof of Theorem \ref{th:3.4}.


\subsubsection{Bazin's theorem}\label{BZS}

The proof of Bazin's theorem for quantum determinants requires two lemmas
of independent interest, which provide
sufficient conditions for certain quantum minors
to commute up to a power of $q$.

Throughout this section, we fix two integers $n < m$, and we consider
quantum minors of the matrix $T$ of generators of $A_q(GL_m)$.
Let $1\le j_1 < \ldots < j_n\le m$. The quantum minor
$\det_q T_{\{1,\ldots , n\},\{j_1,\ldots ,j_n\}}$
is written for short $[j_1,\ldots, j_n]_q$.

\begin{lemma} \label{prop:3.15}
Consider an increasing sequence of integers
$1\leq j_1 < \ldots < j_n < k\leq m$. The following
commutation relation holds for all $i\le n$
$$
[j_1\ldots j_n]_q \, t_{ik} = q^{-1} \, t_{ik}\ [j_1\ldots j_n]_q\ .
$$
\end{lemma}

\Proof We may suppose that $j_s=s$ for $s=1,\ldots ,n$ and $k=n+1$.
The proof is by induction on $n$. For $n=1$, $t_{11}\,t_{12} =
q^{-1}\,t_{12}\,t_{11}$
is one of the defining relations of $A_q(GL_m)$. Assume that the commutation
relation holds for $n-1$. Expand the quantum minor $[1\ldots n]_q$ by its
last row:
$$
[1\ldots n]_q = \sum_{s=1}^n\,(-q)^{s-n}\,t_{ns}\,[1\ldots \hat s \ldots n]_q
\,.
$$
For $i\le n-1$, $t_{i\,n+1}$ commute with $t_{ns}$ and
$$
[1\ldots \hat s \ldots n]_q \, t_{i\,n+1} =
q^{-1}\, t_{i\,n+1}\,[1\ldots \hat s \ldots n]_q
$$
by induction. Therefore $[1\ldots n]_q \, t_{i\,n+1} =
q^{-1}\,t_{i\,n+1}\,[1\ldots n]_q$
for $i\le n-1$. In the remaining case $i=n$, we may use Cayley's
law~\ref{th:3.6}.
Indeed, applying Cayley's law to the relation
$$
[1\ldots \hat s \ldots n]_q \, t_{1\,n+1} =
q^{-1}\, t_{1\,n+1}\,[1\ldots \hat s \ldots n]_q
$$
regarded as an identity between minors of the matrix $T$ of generators of
$A_q(GL_{n+1})$,
we get:
$$
t_{n+1\,n+1}\,\det_q\,T_{\{2,\ldots ,n+1\},\{1,\ldots n\}}
= q\ \det_q\,T_{\{2,\ldots ,n+1\},\{1,\ldots n\}}\,t_{n+1\,n+1}
$$
which is equivalent to
$$
t_{n\,n+1}\,[1\ldots n]_q =q\ [1\ldots n]_q\,t_{n\,n+1}
$$
by translation on the row indices. \cqfd

An immediate corollary of Lemma \ref{prop:3.15} is

\begin{lemma} \label{prop:3.16}
Consider two increasing sequences of integers
$1\leq j_1<\ldots <j_n\leq m$ and $1\leq k_1<\ldots <k_n\leq m$ and suppose
that for some $s\in\{0,\ldots, n\}$, one has $k_s<j_1<j_n<k_{s+1}$. Then,
$$
[j_1\ldots j_n]_q\ [k_1\ldots k_n]_q = q^{2s-n}\,
[k_1\ldots k_n]_q\ [j_1\ldots j_n]_q \ .
$$
\end{lemma}

Thus, for $n=2$, $m=4$, we have
$$
[12]_q\, [34]_q = q^{-2}\, [34]_q\, [12]_q\, , \ \
[14]_q\, [23]_q = [23]_q\, [14]_q\ .
$$

We can now state Bazin's theorem for quantum determinants.

\begin{theorem} \label{th:3.18}
Let $J = \{ j_1 < \ldots < j_n\}$ and
$K = \{ k_1 < \ldots < k_n\}$ be two subsets of $\{1,\ldots m\}$
such that $j_n < k_1$. Then the entries of the matrix $B_n = (b_{st})_{1\leq
s,t \leq n}$
defined by
$$
b_{st} = [j_t, \, (K\!\setminus\, k_s)]_q\ \quad \hbox{for} \ \,
1\leq s,t\leq n \ ,
$$
obey the same commutation rules as the generators of $A_q(GL_n)$ and we have
$$
\det_q \, B_n =
q^{n\choose 2}\ [j_1\ldots j_n]_q\ [k_1\ldots k_n]_q^{n-1}\ .
$$
\end{theorem}

\Proof The proof is by induction on $n\geq2$. For $n = 2$, one can check by
means of Pl\"ucker relations for quantum determinants (described for example
in \cite{TT}) that the entries of
$$
B_2 = \pmatrix{
[j_1 k_2]_q & [j_2 k_2]_q \cr
[j_1 k_1]_q & [j_2 k_1]_q \cr
}
$$
obey the same commutation rules as the generators of $A_q(GL_2)$, and that
$$\det_q \, B_2 = q\ [j_1 j_2]_q\ [k_1 k_2]_q\,.$$
Using Muir's law \ref{th:3.8},
it follows that every $2\!\times\! 2$ submatrix of $B_n$ is a quantum matrix,
and therefore that $B_n$ is itself a quantum $n\!\times\! n$ matrix for every
$n\geq2$. Assume now that
$$
\det_q \, B_{n-1} = q^{n-1\choose 2}\ [j_1\ldots j_{n-1}]_q\
[k_1\ldots k_{n-1}]_q^{n-2}
$$
for all sequences $J$ and $K$ of cardinality $n\!-\! 1$ satisfying the
hypothesis of Theorem \ref{th:3.18}. From Theorem \ref{th:3.2}, it
results that
$$
\det_q\, B_n = |B_n|_{nn}\ \det_q\, B_n^{nn}\ .
$$
Now Muir's law and the induction hypothesis show that
$$
\det_q \, B_n^{nn} = q^{n-1\choose 2}\
[j_1 j_2 \ldots j_{n-1} k_n]_q\ [k_1 k_2\ldots k_n]_q^{n-2}\ .
$$
On the other hand, expanding all entries of $|B_n|_{nn}$ according to
Theorem \ref{th:3.2}, and applying
Bazin's theorem for quasi-determinants, one obtains
$$
|B_n|_{nn} = [k_1 \ldots k_n]_q\ [j_1 j_2 \ldots j_{n-1} k_n]_q^{-1}\
[j_1 \ldots j_n]_q\ .
$$
The claim follows now from Muir's law, which shows that
$$
[j_1 \ldots j_n]_q\ [j_1 j_2\ldots j_{n-1} k_n]_q
= q^{-1}\
[j_1 j_2 \ldots j_{n-1} k_n]_q\ [j_1 \ldots j_n]_q \ ,
$$
and from Proposition \ref{prop:3.16}. \cqfd

As an illustration, take $n=3$, $J=\{1,2,3\}$ and $K=\{4,5,6\}$.
Then, Bazin's theorem reads
$$
\left|\matrix{
[145]_q & [245]_q & [345]_q \cr
[146]_q & [246]_q & [346]_q \cr
[156]_q & [256]_q & [356]_q \cr
}\right|_q
= q^3\ [123]_q\ [456]_q^2
\ .
$$
%


\subsubsection{Schweins' series}

Using Theorem \ref{th:3.2}, one readily deduces from Schweins' series
for quasi-determinants (Theorems \ref{th:2.27}, \ref{th:2.28})
the following quantum analogues. Here again identities are stated for
quantum determinants of order 3 and 4 only, the general case being easily
understood from these. The notations for quantum minors are those
introduced in Section\ref{BZS}.

\begin{theorem} \label{th:3.20}
The maximal minors of a $3\times 6$ quantum matrix
satisfy the relation
$$
[123]_q^{-1} [124]_q
=
[123]_q^{-1} [125]_q [235]_q^{-1} [234]_q +
[253]_q^{-1} [256]_q [356]_q^{-1} [354]_q +
[563]_q^{-1} [564]_q\ .
$$
The maximal minors of a $4\times 8$ quantum matrix satisfy the relation
$$
[1234]_q^{-1} [1235]_q
=
[1234]_q^{-1} [1236]_q [2346]_q^{-1}
[2345]_q
$$
$$
+[2364]_q^{-1} [2367]_q [3467]_q^{-1} [3465]_q
+[3674]_q^{-1} [3678]_q [4678]_q^{-1} [4675]_q
+[6784]_q^{-1} [6785]_q\ .
$$
\end{theorem}

\begin{theorem} \label{th:3.21}
The minors of a $3\times 4$ quantum matrix satisfy the relation
$$
\left|\matrix{
t_{11} & t_{12} & t_{13}\cr
t_{21} & t_{22} & t_{23}\cr
t_{31} & t_{32} & t_{33}\cr
}\right|_q^{-1}
\left|\matrix{
t_{11} & t_{12} & t_{14}\cr
t_{21} & t_{22} & t_{24}\cr
t_{31} & t_{32} & t_{34}\cr
}\right|_q
$$
$$
= \,
\left|\matrix{
t_{11} & t_{12} & t_{13}\cr
t_{21} & t_{22} & t_{23}\cr
t_{31} & t_{32} & t_{33}\cr
}\right|_q^{-1}
\left|\matrix{
t_{11} & t_{12} \cr
t_{21} & t_{22} \cr
}\right|_q
\left|\matrix{
t_{11} & t_{13} & t_{14}\cr
t_{21} & t_{23} & t_{24}\cr
t_{31} & t_{33} & t_{34}\cr
}\right|_q
\left|\matrix{
t_{11} & t_{13} \cr
t_{21} & t_{23} \cr
}\right|_q^{-1}
$$
$$
+
\left|\matrix{
t_{11} & t_{13}\cr
t_{21} & t_{23}\cr
}\right|_q^{-1}
t_{11}
\left|\matrix{
t_{13} & t_{14}\cr
t_{23} & t_{24}\cr
}\right|_q
t_{13}^{-1}
$$
$$
+\,
t_{13}^{-1} t_{14}
\ .
$$
The minors of a $4\times 5$ quantum matrix satisfy the relation
$$
\left|\matrix{
t_{11} & t_{12} & t_{13} & t_{14}\cr
t_{21} & t_{22} & t_{23} & t_{24}\cr
t_{31} & t_{32} & t_{33} & t_{34}\cr
t_{41} & t_{42} & t_{43} & t_{44}\cr
}\right|_q^{-1}
\left|\matrix{
t_{11} & t_{12} & t_{13} & t_{15}\cr
t_{21} & t_{22} & t_{23} & t_{25}\cr
t_{31} & t_{32} & t_{33} & t_{35}\cr
t_{41} & t_{42} & t_{43} & t_{45}\cr
}\right|_q
$$
$$
= \,
\left|\matrix{
t_{11} & t_{12} & t_{13} & t_{14} \cr
t_{21} & t_{22} & t_{23} & t_{24} \cr
t_{31} & t_{32} & t_{33} & t_{34} \cr
t_{41} & t_{42} & t_{43} & t_{44} \cr
}\right|_q^{-1}
\left|\matrix{
t_{11} & t_{12} & t_{13} \cr
t_{21} & t_{22} & t_{23} \cr
t_{31} & t_{32} & t_{33} \cr
}\right|_q
\left|\matrix{
t_{11} & t_{12} & t_{14} & t_{15}\cr
t_{21} & t_{22} & t_{24} & t_{25}\cr
t_{31} & t_{32} & t_{34} & t_{35}\cr
t_{41} & t_{42} & t_{44} & t_{45}\cr
}\right|_q
\left|\matrix{
t_{11} & t_{12} & t_{14} \cr
t_{21} & t_{22} & t_{24} \cr
t_{31} & t_{32} & t_{34} \cr
}\right|_q^{-1}
$$
$$
+
\left|\matrix{
t_{11} & t_{12} & t_{14} \cr
t_{21} & t_{22} & t_{24} \cr
t_{31} & t_{32} & t_{34} \cr
}\right|_q^{-1}
\left|\matrix{
t_{11} & t_{12} \cr
t_{21} & t_{22} \cr
}\right|_q
\left|\matrix{
t_{11} & t_{14} & t_{15} \cr
t_{21} & t_{24} & t_{25} \cr
t_{31} & t_{34} & t_{35} \cr
}\right|_q
\left|\matrix{
t_{11} & t_{14} \cr
t_{21} & t_{24} \cr
}\right|_q^{-1}
$$
$$
+
\left|\matrix{
t_{11} & t_{14}\cr
t_{21} & t_{24}\cr
}\right|_q^{-1}
t_{11}
\left|\matrix{
t_{14} & t_{15}\cr
t_{24} & t_{25}\cr
}\right|_q
t_{14}^{-1}
$$
$$
+\,
t_{14}^{-1} t_{15}
\ .
$$
\end{theorem}






\end{document}